\documentclass[useAMS,amsmath,myaasmacros,amssymb,epsfig,usegraphicx,twocolumn,usenatbib]{mn2e}
\def\aj{AJ}
\def\araa{ARA\&A}
\def\apj{ApJ}
\def\apjl{ApJ}
\def\aap{A\&A}
\def\mnras{MNRAS}
\def\nat{Nature}
\def\na{NewA}

\def\Plus{\texttt{+}}
\def\Minus{\texttt{-}}

\usepackage{amsmath}

\newif\ifAMStwofonts

\makeatother

\title[Spinning dark matter halos]
{Spinning dark matter halos promote bar formation}

\author[Saha \& Naab]
{Kanak Saha$^{1}$\thanks{E-mail:saha@mpe.mpg.de} \& Thorsten Naab$^{2}$\\
$^{1}$Max-Planck-Institut f\"ur Extraterrestrische Physik, Giessenbachstrasse, D-85748
Garching, Germany\\ $^{2}$Max-Planck Institut f\"ur Astrophysik, Karl-Schwarzschild Straße 1, 85748 Garching, Germany}
 
\begin{document}

\date{Accepted xxxx Month xx. Received xxxx Month xx; in original form
2013 April 2} 
\pagerange{\pageref{firstpage}--\pageref{lastpage}} \pubyear{2013}
\maketitle

\label{firstpage}

\begin{abstract}
Stellar bars are the most common non-axisymmetric structures in
galaxies and their impact on the evolution of disc galaxies at
all cosmological times can be significant. Classical theory predicts
that stellar discs are stabilized against bar formation if embedded in
massive spheroidal dark matter halos. However, dark matter halos have
been shown to facilitate the growth of bars through resonant
gravitational interaction. Still, it remains unclear why some
galaxies are barred and some are not. In this study, we demonstrate that
co-rotating (i.e., in the same sense as the disc rotating) dark matter halos 
with spin parameters in the range of 
$0 \le \lambda_{\mathrm{dm}} \le 0.07$ - which are a definite 
prediction of modern cosmological models - promote the formation of bars 
and boxy bulges and therefore can play an important role in the formation of
pseudobulges in a kinematically hot dark matter dominated disc
galaxies. We find continuous trends for models with higher halo spins:
bars form more rapidly, the forming slow bars are stronger, and
the final bars are longer. After $2$~Gyrs of evolution, the amplitude
of the bar mode in a model with $\lambda_{\mathrm{dm}} =  0.05$ is a
factor of $\sim 6$ times higher, $A_2/A_0 = 0.23$, than in the
non-rotating halo model. After $5$~Gyrs, the bar is $\sim$ 2.5 times
longer. The origin of this trend is that more rapidly spinning
(co-rotating) halos provide a larger fraction of trailing dark matter
particles that lag behind the disc bar and help growing the bar by
taking away its angular momentum by resonant interactions. A
counter-rotating halo suppresses the formation of a bar in our
models. We discuss potential consequences
for forming galaxies  at high-redshift and present day low mass
galaxies which have converted only a small fraction of their baryons
into stars.     
 
\end{abstract}

\begin{keywords}
galaxies:halos -- galaxies: spiral -- galaxies: structure -- galaxies: kinematics and 
dynamics -- galaxies: evolution
\end{keywords}

\section{Introduction}
\label{sec:introduc}

Optical and near-infrared observations indicate that nearly $2/3$ of
 disc galaxies in the local universe are strongly barred
\citep{Eskridgeetal2000,Grosboletal2004,MenendezDelmestreetal2007,Barazzaetal2008}. 
However, the fraction of barred
disc galaxies might have significantly increased from high redshift
until the present day. Whereas some studies indicate that it is nearly 
constant up to $z \sim 1$ \citep{Jogeeetal2004,Elmegreenetal2004,MarinovaJogee2007}
other studies indicate a decrease in bar fraction towards higher
redshift to values as low as $20$\% at $z \sim 0.84$
\citep{Shethetal2008}. A common message from these observational
studies is that bars can form in their host stellar discs, embedded in
diverse environments, and survive through $7 - 8$~Gyrs of look-back
time. 
\footnote{Throughout the text, by co-rotating dark matter halo, we refer 
to a halo rotating in the same direction as the embedded disc and opposite for
the counter-rotating case.} 

The formation mechanisms of a bar have been mostly studied with N-body
simulations. They can either form as a result of swing-amplification 
of gravitationally unstable $m=2$ modes in a stellar disc
\citep{Toomre1964,Goldreich-Tremaine1979,Toomre1981,CombesSanders1981,
SellwoodWilkinson1993,Polyachenko2013}, triggered by galaxy interactions
 and mergers \citep{Noguchi1987,BarnesHerquist1991,MiwaNoguchi1998} or 
through the cooperation of orbital streams which eventually lead to formation 
of a bar \citep{Earn-LyndenBell1996}. 
Interaction with dark matter halo substructures \citep{Romano-Diazetal2008} 
as well as the resonant gravitational interaction with the surrounding dark 
matter 'particles' has been shown to promote bar formation 
\citep{Athana2002, Dubinskietal2009}. 
Without interactions or in otherwise isolated environments, the
formation and evolution of bars in disc galaxies strongly depends on
the properties of the surrounding dark matter halo. It has been known
for a long time that a dynamically hot halo component is required to
stabilize an otherwise gravitationally unstable rotating
self-gravitating disc against bar formation. 
According to \cite{Ostriker-Peebles1973}, a stellar disc will go bar
unstable if the ratio of total rotational kinetic energy to the potential energy
exceeds $0.14 \pm 0.03$. If the disc is embedded in a 
non-rotating dark matter halo, the potential energy increases and the
disc is stabilized against bar formation \citep{Hohl1976}. This finding was 
again confirmed by \cite{ELN1982}
and quantified in terms of observable parameters like disc scale
length and circular velocity. However, a study by \cite{Athanamisi2002,Athanassoula2003}
indicated that strong bars can form in cold discs embedded in massive dark  
matter halos casting doubts on the general applicability of the above
criteria \citep{Athanassoula2008}.

Traditionally, the dark matter halo was treated as a rigid component which 
does not respond to the dynamics of the disc stars. However, 'live' dark matter 
halos enhance the growth of bars in model galaxies by resonant interactions 
of disc stars with halo 'particles' taking away the disc's angular momentum
\citep{DebattistaSellwood1998,Athana2002, Athanassoula2003,Holley-Bockelmannetal2005, 
WeinbergKatz2007a,Ceverinoklypin2007,Sahaetal2012}. The exchange of angular momentum
between the disc and halo has been a subject of active research over
the last two decades and our understanding of the bar instability has
improved significantly. 

Most studies focused on the formation of bars in kinematically cold
discs (which easily form bars) embedded in close to spherical
non-rotating dark matter halos. The evolution of systems, especially  
dynamically hot discs which might even be dark matter dominated was
left unexplored, mainly because such systems were not expected to form
bars at all. However, such systems exist. Low Surface Brightness (LSB) 
galaxies are generally isolated, without nearby companions and are 
dominated by dark matter at all radii \citep{deBloketal2001}. The evolution 
of such galaxies is expected to depend entirely on internal processes and it 
remains unclear how their discs become bar unstable 
\citep{MatthewsGallagher1997,Pohlenetal2003}, as $\sim 30\%$ of LSB galaxies 
reported in the sample of \cite{MatthewsGallagher1997} are barred.  
The study of \cite{Sahaetal2013} indicates, that
hot stellar discs inside a massive non-rotating halo are stable against
the formation of strong bars; they can form only weak bars over
several Gyrs. In this paper, we test the hypothesis that halo rotation
- in accordance with standard cosmological models - can promote the
formation of strong bars even in the presence of massive dark matter halos.

One of the definite predictions from large scale dark matter
simulations with $\Lambda$CDM cosmological models is that dark matter halos
acquire a certain amount of spin due to the cosmic shear and mergers
\citep{Steinmetz1995,Bullocketal2001,Vitvitskaetal2002,Springeletal2005,
Bettetal2007} and they are hardly
non-rotating. Their spin distribution peaks at a value $\lambda_{\mathrm{dm}} \sim
0.035$ with a tail towards halos with even higher $\lambda_{\mathrm{dm}}$-values. It
has been suggested that rotating halos could enhance the growth of a
bar \citep{Weinberg1985} as these halos have a higher fraction of dark
matter particles on prograde orbits which can absorb disc angular
momentum even more efficiently. We conduct a systematic investigation
of the effect of halo spin on the evolution of a hot stellar disc
embedded in a massive dark matter halo. The effect is tested with a galaxy model
which is expected to be stable (or weakly unstable) against bar formation
according to most of the conventional criteria of bar instability.
We find that the spin of a dark matter halo has substantial effect on the
bar instability and thereby on the galaxy evolution and carry out a
detailed investigation using - cosmologically motivated - spinning dark matter halos.  

\begin{figure}
\rotatebox{270}{\includegraphics[height=8.5 cm]{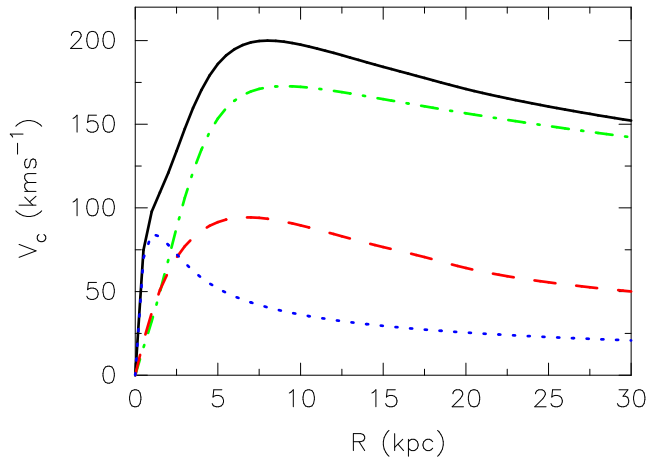}}
\caption{Circular velocity of disc stars (red dashed), bulge stars
  (blue dotted) and dark matter (green dot-dashed) of our fiducial
  initial condition galaxy. The galaxy disc dominated by dark matter
  (or a spheroidal bulge component) at all radii.} 
\label{fig:vcdbh}
\end{figure}

The paper is organized as follows. Section~\ref{sec:modelsetup} 
describes the construction of initial galaxy models with different halo 
spins and the basics of the simulation set up. Section~\ref{sec:instability} 
outlines different conventional criteria of bar instability. Bar evolution 
and the morphology of the disc is presented in section~\ref{sec:evol}. 
Section~\ref{sec:OPspin} and section~\ref{sec:bulge} discuss the effect of 
halo spin on the Ostriker-Peebles criteria and bulge formation respectively.
Section~\ref{sec:TAM} is devoted to angular momentum transfer and density 
wakes in the halo. Section~\ref{sec:reso} describes the resonances in the 
system. The discussion and conclusions are presented in section~\ref{sec:discuss}.
 
\begin{figure}
\rotatebox{270}{\includegraphics[height=8.5 cm]{ToomreQvsRkpc-ABCDE-12June13.ps}}
\caption{Radial profile of the Toomre Q parameter of the stellar
  disc. The dashed black line shows the initial condition at t=0. The
  other lines show the profiles after 2 Gyrs evolution for model A+00
  (black solid), B+03 (green), C+05 (blue), D+07 (red), and the model
  with the counter-rotating halo E-05 (cyan). At any radius and at any
  time the stellar disc is stable in terms of Toomre $Q > 1.$}
\label{fig:toomreQ}
\end{figure}

\begin{figure*}
\rotatebox{0}{\includegraphics[height=19.0cm]{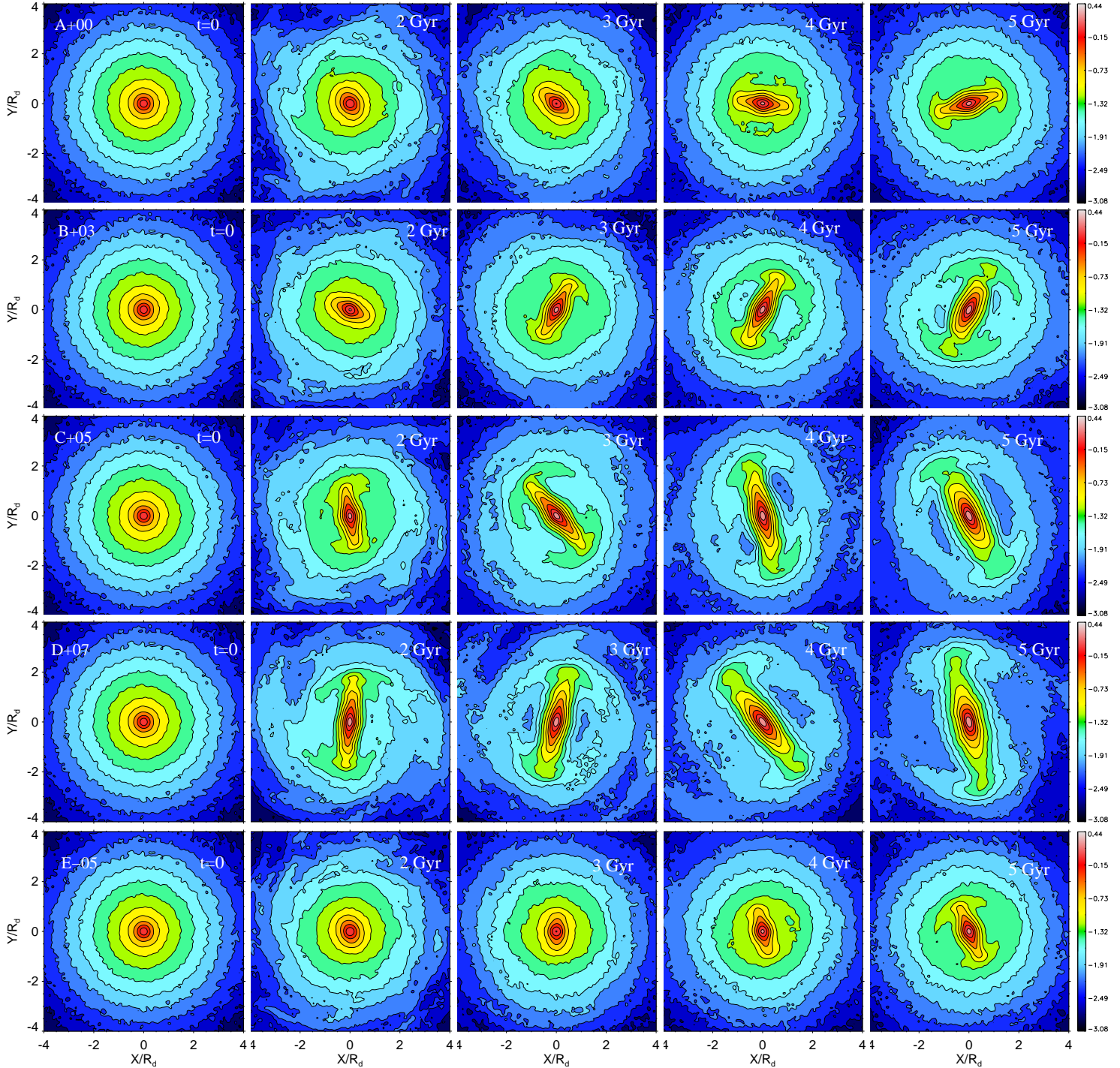}}
\caption{Time evolution (from left to right at t = 0, 2, 3, 4, 5 Gyrs)
  of the stellar surface density profiles for model with varying dark
  matter halo spins. From top to bottom the spin of the co-rotating
  (with the stellar disc) dark matter halos increases (models A+00,
  B+03, C+05, D+07). The bottom row shows the evolution of a model 
  but with a counter-rotating halo (E-05). Models with more rapidly 
   spinning co-rotating dark halos
  become bar unstable earlier and develop longer
  bars. Counter-rotating halos do not promote bar formation.}   
\label{fig:fdenmap}
\end{figure*}

\section{galaxy models with spinning halo}
\label{sec:modelsetup}
We construct $5$ equilibrium models of galaxy with initially spinning
dark matter halos (hereafter, ISDM) using the self-consistent method
of \citet{KD1995}. Each galaxy model consists of a live disc, a dark
matter halo, and a classical bulge. Flattened models of ISDMs are
constructed by reversing the velocities of halo particles with
negative angular momenta perpendicular to the disc plane, which
remains a valid solution of the collisionless Boltzmann equation
\citep{Lynden-Bell62, Hernquist1993,SahaGerhard2013b}. In this way, 
we have constructed one model
with a non-rotating halo (A+00), three models with co-rotating halos
with spin parameters $\lambda_{\mathrm{dm}} \sim$ 0.03, 0.05, and 0.07 (B+03, C+05,
D+07) and one model with a counter-rotating halo (E-05) with a spin
parameter of $\lambda_{\mathrm{dm}} \sim$ 0.05, similar to model C+05 but
the halo is rotating in the opposite direction of the disc. For the
definition of the spin parameter ($\lambda_{\mathrm{dm}}$) of the dark
matter halos we follow \citep{Bullocketal2001},

\begin{equation}
\lambda_{\mathrm{dm}} \equiv \frac{1}{\sqrt{2}} \frac{\sum_{i}{m_i \vec{r_i} \times
\vec{v_i}}}{M_{vir}V_{vir}R_{vir}},
\end{equation}

\noindent where $m_i$, $\vec{r_i}$ and $\vec{v_i}$ are the mass, radius and
velocity of a dark matter particle; $V_{\mathrm{vir}}^2 = G
M_{\mathrm{vir}}/R_{\mathrm{vir}}$ denotes the circular velocity at
the virial radius ($R_{\mathrm{vir}}$). The values of
$\lambda_{\mathrm{dm}}$ are shown in Table~\ref{tab:paratab} and reasonably cover
the range of spin distributions of dark matter halos from
cosmological simulations. An analysis of the Millennium simulation
\citep{Springeletal2005} indicates that the halo spin 
parameter only weakly depends on halo mass and shape
\citep{Bettetal2007}. Therefore, we changed only the halo
spin parameter keeping its mass and shape unchanged to single-out the
effect of halo angular momentum on the disc evolution.     

\begin{figure}
\rotatebox{270}{\includegraphics[height=8.5 cm]{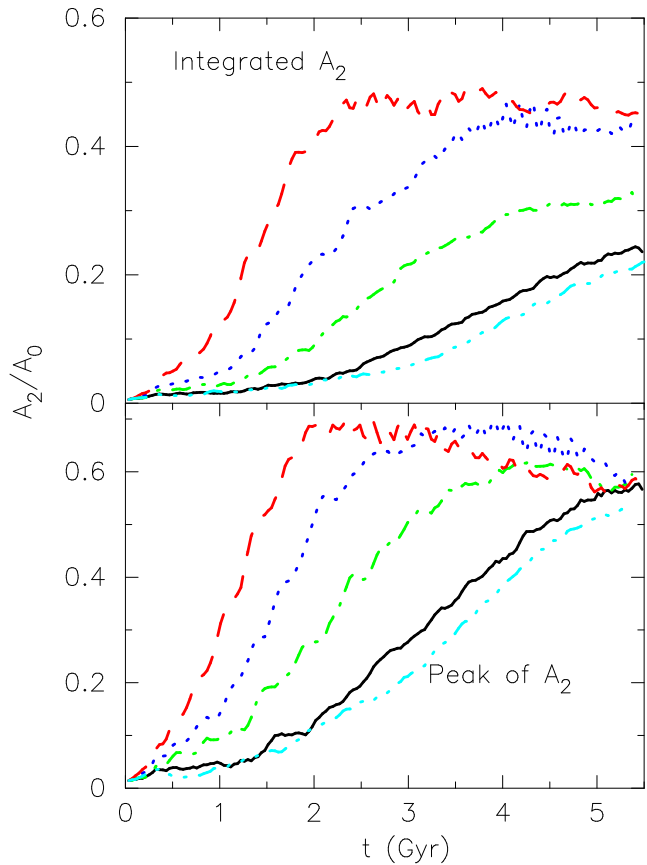}}
\caption{Time evolution of the peak of the $m=2$ Fourier amplitude of the surface
  density. Red dashed line indicates D+07, blue dotted C+05, green dash-dot B+03, 
black solid line A+00, and cyan dash-dot-dot E-05. There is a clear trend for models 
with more rapidly spinning 
co-rotating halos to develop stronger bars at earlier evolutionary phases (e.g., D+07).}
\label{fig:A2ps}
\end{figure}

The disc has an exponentially declining surface density with a
scale-length $R_d$, a constant scale-height $h_z$ and mass $M_d$. The
outer radius of the disc is truncated at $7.0 R_d$ with a truncation
width of $0.5 R_d$ within which the stellar density smoothly drops to
zero. The live dark matter halo is modelled with a distribution function 
given by the lowered Evans model \citep{Evans1993} which has a constant density 
core and produce a nearly flat rotation curve in the outer parts. The initial flattening of 
the halo is $q  = 0.8$; the mass and tidal radius of the halo are given by 
$M_h = 10.32 M_d$ and $R_h = 36 R_d$ respectively.  
The preexisting classical bulge is represented by a King
distribution function \citep{King1966}. The mass and tidal radius of the bulge 
are $M_b =0.18 M_d$ and $R_b = 3.2 R_d$ respectively. For relevant details on 
model construction, the reader is referred to \cite{Sahaetal2010, Sahaetal2012}. 

\begin{figure}
\rotatebox{270}{\includegraphics[height=8.5 cm]{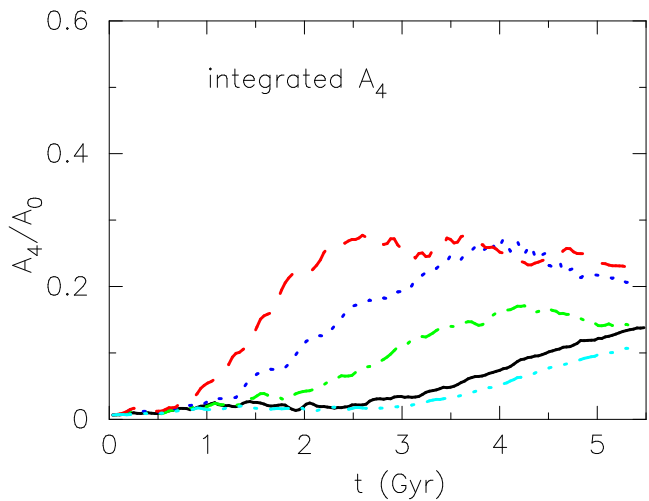}}
\caption{Same as in Fig.~\ref{fig:A2ps} but for the integrated $m=4$ Fourier amplitude
of the surface density.}
\label{fig:A4ps}
\end{figure}

In terms of astronomical units, $M_d = 1.7 \times 10^{10} M_{\odot}$
and $R_d =3.0$ kpc, so that the time unit is $24.0$ Myr. 
The initial value of $h_z = 300$~pc. The total
mass of the dark matter halo and the bulge is $M_h = 1.73 \times
10^{11} M_{\odot}$ and $M_b = 3.0 \times 10^{9} M_{\odot}$,
respectively.  We use a total of $3.5 \times 10^6$ particles to simulate 
each model; assigning $1.2\times10^6$ to the disc, $1.8\times10^6$ to the 
halo, and $0.5\times10^6$ to the bulge. The softening lengths for
the disc, bulge and halo are $9.0$, $5.9$ and $23.6$~pc, respectively,
following \cite{McMillan2007}. All simulations were performed using 
the Gadget code \citep{Springeletal2001} with an opening angle of 
$\theta =0.7$ for the tree code performance and evolved all the models
for about $5.5$~Gyr. The half-mass rotation period (dynamical time-scale) 
at $2.5 R_d$ is $237$~Myr ($37.75$~Myr). At the end of the simulation 
the total energy was conserved to about $\sim 0.03\%$.

In Fig.~\ref{fig:vcdbh}, we show the circular velocity curve for the
galaxy model constructed as described above. The model is dominated by
dark matter (and a spheroidal bulge component) right from the central
region, similar to some giant LSB galaxies \citep{deBloketal2001,Lellietal2010}. 
The initial stellar disc is kinematically hot and stable against local gravitational
instability as can be seen from the initial radial profile of the
Toomre Q \citep{Toomre1964} parameter (see Fig.~\ref{fig:toomreQ}), defined as
$Q(R)={\sigma_r(R) \kappa(R)}/{3.36 G \Sigma(R)}$, where $\sigma_r$ is
the radial velocity dispersion, $\kappa$ is the epicyclic frequency
and $\Sigma$ is the surface density of stars.

\begin{table}
\caption[ ]{Initial spin of the dark matter halos, values of Ostriker-Peebles parameter
 and properties of bar at $5$~Gyr. The unit of pattern speed $\Omega_{\mathrm{B}}$ is in 
km s$^{-1}$kpc$^{-1}$.}
\begin{flushleft}
\begin{tabular}{cccccc}  \hline\hline 
Models     &$\lambda_{\mathrm{dm}}$ & $t_{\mathrm{OP}}$& $R_{\mathrm{bar}}/R_d$ & $R_{\mathrm{cr}}/R_d$ & $\Omega_{\mathrm{B}}$ \\
       & &  &  & &\\
\hline
\hline
A\Plus00   & 0.000 & 0.039 & 1.32  & 3.28 & 18.3\\
B\Plus03   & 0.032 & 0.053 & 1.61  &3.91 & 15.0\\
C\Plus05   & 0.053 & 0.078 & 1.78  &4.72 & 12.2\\
D\Plus07   & 0.074 & 0.115 & 2.60  &4.76 & 12.0\\
E\Minus05   &-0.053 & 0.078 &1.07 & 3.32 & 18.0\\

\hline
\end{tabular}
\end{flushleft}
\footnote{ABC:}
{$t_{\mathrm{OP}}$ denotes the Ostriker-Peebles criteria explained in section~3. 
$R_{\mathrm{bar}}$ is the bar size, $R_{\mathrm{cr}}$ the co-rotation radius and 
$\Omega_{\mathrm{B}}$ is the bar pattern speed.}
\label{tab:paratab}
\end{table}

\section{criteria for bar instability}
\label{sec:instability}
Many simulations have shown that a strong bar forms easily in a cold, rotation
dominated stellar disc 
\citep[and references therein]{Hohl1971,SellwoodWilkinson1993,Sahaetal2012, Athanassoula2012}. 
Based on the study of the stability of Maclaurin discs \citep{BT1987}
and N-body simulations of galactic discs, \cite{Ostriker-Peebles1973}
suggested that a stellar disc would become bar unstable if the ratio
$t_{\mathrm{OP}} = T_{\mathrm{rot}}/|W|$ exceeds a critical value of
$0.14 \pm 0.03$. In our simulations, we compute the Ostriker-Peebles
criterion for each model in a slightly different way. Assuming the
virial theorem applies to the isolated galaxy, it can be shown that \citep{BT1987}  
\begin{eqnarray}
t_{\mathrm{OP}} = 1 / (2 + T_{\mathrm{rand}}/T_{\mathrm{mean}}).
\end{eqnarray}

\begin{eqnarray}
T_{\mathrm{rand}} = \frac{1}{2}\int{(v - \bar{v})^2 \rho(x) d^3 x} 
\end{eqnarray} 
is the kinetic energy in random motions and 
\begin{eqnarray}
T_{\mathrm{mean}} = \frac{1}{2}\int{{\bar{v}}^2 \rho(x) d^3 x}
\end{eqnarray}
denotes the kinetic energy in the mean (primarily rotational)
motion. To compute $t_{\mathrm{OP}}$ for each model galaxy, we need to
add the contribution to  the total kinetic energy budget from all the
sub-components in the model e.g., bulge (B), disc (D)  and halo (H), so that
\begin{eqnarray}
T_{\mathrm{rand}} = T_{\mathrm{rand,B}} + T_{\mathrm{rand,D}} + T_{\mathrm{rand,H}},
\end{eqnarray}
and
\begin{eqnarray}
T_{\mathrm{mean}} = T_{\mathrm{mean, B}} + T_{\mathrm{mean, D}} + T_{\mathrm{mean, H}}.
\end{eqnarray}

\begin{figure}
\rotatebox{270}{\includegraphics[height=8.5 cm]{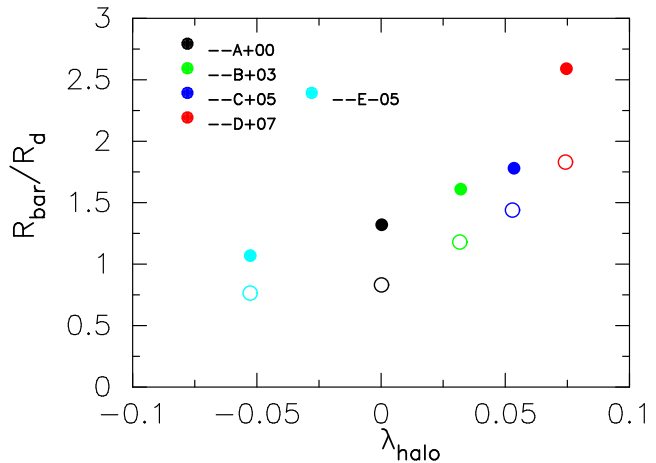}}
\caption{Bar size as a function of halo spin. Open circles indicate
the sizes of the bars at $T = 3$~Gyr and filled circles at $T = 5$~Gyr.
Open circle of a particular color refers to the same model as denoted by the 
filled circle of the same color. Bars in models with larger halo spin are 
larger at all times.}
\label{fig:barsize}
\end{figure}

\begin{figure}
\rotatebox{270}{\includegraphics[height=8.5 cm]{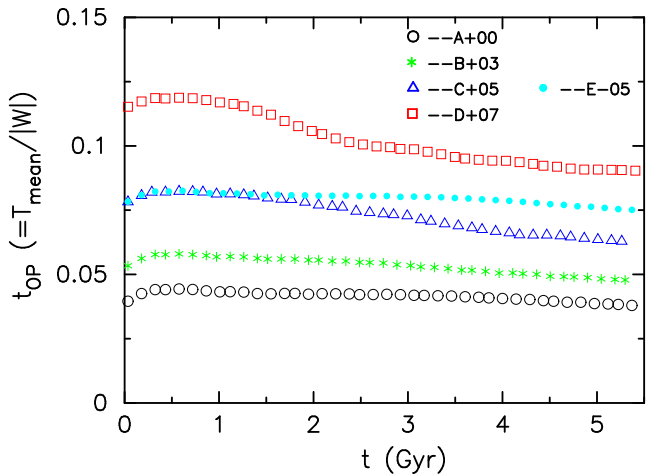}}
\caption{Time evolution of the Ostriker-Peebles stability parameter,
  $t_{\mathrm{OP}}$, (section \ref{sec:instability}) for models A+00 to
  E-05. According to this criterion all models should be stable
  against bar formation. However, this criterion accounts for the fact
that the models with halo co-rotation are more prone to instabilities
(e.g. higher values for $t_{\mathrm{OP}}$).}
\label{fig:top}
\end{figure}

Since the initial bulge and disc parameters are identical for all the
models in our simulation, the variation in the initial
$t_{\mathrm{OP}}$ arise due to the variation only in the halo spin. It is
interesting to note that the total kinetic energy in a galaxy is
independent of the sense in which particles are moving. So the total kinetic
energy in a halo with a fixed spin parameter is the same, independent
of whether it is co-rotating or counter-rotating with respect to the
disc; $t_{\mathrm{OP}}$ is, however, sensitive to the net amount of halo
rotation. It is worth noting that D+07 has a higher value than
other models and is likely to be unstable according to Ostriker-Peebles criteria. 
We will return to this issue in a later section. 

A similar global criterion, suggested by \cite{ELN1982}, states that if
$\alpha_{\mathrm{eln}}=V_{\mathrm{peak}}/\sqrt{G M_d/R_d} < 1.1$, a self-gravitating stellar
disc would become bar unstable. Here, $V_{\mathrm{peak}}$ refers to the maximum of the 
rotational velocity. Basically, the dimensionless parameter 
$\alpha_{\mathrm{eln}}$ measures the ratio of the total mass within a given radius
to that in the stellar disc, such that a massive halo would provide stability 
against bar formation. For a bare exponential disc without any bulge and dark halo, the
value of $\alpha_{\mathrm{eln}} = 0.63$ \citep{ELN1982} and would be subject to vigorous 
bar instability. For our initial conditions $\alpha_{\mathrm{eln}} = 1.3$
indicating clearly that all the galaxy models would be stable against
bar formation. Note that the value of $\alpha_{\mathrm{eln}}$ is insensitive to  
the halo rotation. 

According to \cite{Toomre1981} and a number of other studies
\citep[and references therein]{SellwoodWilkinson1993,Dubinskietal2009}, 
bars in N-body simulations 
are formed through the swing-amplification mechanism. For this
mechanism to work, the disc needs to have no inner Lindblad
resonance (ILR), allowing the feedback loop to be in action. It has been
shown that to achieve strong swing amplification of an  $m=2$ bar mode, a
stellar disc requires to have $Q < 2$ and $1 < X_2 < 3$
\citep{JulianToomre1966,Toomre1981}.  
Here $X_2(R) = R \kappa^2(R)/(4\pi G \Sigma(R))$ is the swing-amplification parameter 
 for an $m=2$ mode in the tight winding approximation \citep{Goldreich-Tremaine1979}.
For the models considered here, the value of $X_2 = 6.4$ at $2.2 R_d$ which marks 
the location of the peak of the disc's circular velocity. According to all classical
stability criteria, the stellar discs in our simulation should be linearly stable 
(or weakly unstable) to strong bar formation.

\section{Bar evolution - morphology}
\label{sec:evol}
In Fig.~\ref{fig:fdenmap}, we show time evolution of the face-on surface
density maps (from left to right) of the initially axisymmetric
stellar disc embedded in a massive dark matter halo with different
spin parameters (from top to bottom, see Table~\ref{tab:paratab}). The
model with a  non-rotating halo (A+00) remains nearly axisymmetric for
about $2$~Gyrs; however, as the spin of the halo increases, the disc
starts forming a strong bar even earlier. It is clearly visible that
the model with the strongest halo rotation (D+07) develops the strongest
bar. The model with the counter-rotating halo (E-05) develops a bar in a 
similar way to the non-rotating model (A+00).  

We have quantified this finding in Fig.~\ref{fig:A2ps} which shows the
time evolution of the $m=2$ Fourier component of the surface density
normalized by the axisymmetric ($m=0$) component, $A_2 / A_0$ for all
the models. In the lower panel of Fig.~\ref{fig:A2ps}, we have shown
the time evolution of the peak of the $A_2 / A_0$ values -- which is
a valid definition of bar amplitude used frequently in literatures 
\citep[and references therein]{Sahaetal2010,Sahaetal2013}. Till about
$\sim 3$~Gyr, there is a clear correlation between bar amplitude
and the halo spin -- higher spin leads to higher $A_2 / A_0$. However,
beyond $3$~Gyr, no such correlation is evident from the figure; while
all the bar amplitudes nearly approaching to a single value at $\sim 5$~Gyr 
but with different bar sizes (see Fig.~\ref{fig:fdenmap}, section~\ref{sec:size} and 
Table~\ref{tab:paratab}). 
To avoid such intriguing situation, we compute the bar amplitude as integrated 
contribution of the $m=2$ Fourier component over a radial range (which takes 
into account the radial variation of $A_2 / A_0$ in the disc) as 

\begin{equation}
\tilde{\frac{A_2}{A_0}} =\frac{1}{R_{\mathrm{max}}} \int_{0}^{R_{\mathrm{max}}}{\frac{A_2}{A_0}(R) dR},
\end{equation}

\noindent where we chose $R_{\mathrm{max}} = 2.5 R_d$, the region which confines 
most of the bars in our galaxy models at all times. The upper panel of Fig.~\ref{fig:A2ps}, 
shows the integrated $\tilde{\frac{A_2}{A_0}}$ as a function of time. Within
$3$~Gyr, both methods give approximately same time variation for the bar 
amplitude, although they are quantitatively different. But after $3$~Gyr, these two
methods give entirely different result in that the integrated 
$\tilde{\frac{A_2}{A_0}}$ brings out the positive correlation with halo spin while 
the peak of $A_2/A_0$ does not. 
At about $2$~Gyr, the integrated amplitude of the stellar bar in the halo with 
$\lambda_{dm} = 0.074$ is higher by a factor of $\sim 11$ compared to the 
non-rotating halo case. Both panels of Fig.~\ref{fig:A2ps} suggest that there 
is a clear trend -- a more rapidly spinning co-rotating dark halo promotes 
the early formation of a strong bar.  

From Fig.~\ref{fig:fdenmap}, a systematic increase in the bar length is
noticeable as the spin of the co-rotating halo increases. A large-scale 
bar is formed in the model D+07 after $2$~Gyr and by the end of
the simulation at $5$~Gyr the bar extends to $\sim 8$~kpc ($\sim$ the peak of the 
rotation curve, see Fig.~\ref{fig:vcdbh}). Interestingly, such a long 
and strong bar turns out to be a slow bar since it has formed and remain so throughout 
the simulation; as discussed, in detail, in section~\ref{sec:reso}, all the bars in our 
simulations remain as slow bars at all times. Swing amplification \citep{Toomre1981}
might have played an important role in the formation and growth of such a bar as
our preliminary check showed no ILR (Inner Lindblad Resonances) which is crucial
for establishing the feedback loop in the stellar disc. However, a 
detailed prediction of the initial bar pattern speed from most bar formation theories 
remains unclear.   
Note that in most models such
bars do not trigger any two-armed spiral feature, except transient
spirals; primarily because of the higher values of Toomre Q and $X_2$ parameter of
the disc. As a bar grows, the combined effect of the bar and transient
spirals heat the stars throughout the disc \citep{JenkinsBinney1990,Sahaetal2010} and
as a result of that the Toomre Q of the disc rises to even higher values
(see Fig.~\ref{fig:toomreQ}). This, in turn, reduces the possibility
of growing spirals through swing amplification.        

Interestingly, the growth of a stellar bar in our counter-rotating
halo model (E-05) is significantly suppressed compared to the
co-rotating model with the same spin (C+05). Substantial differences
between C+05 and E-05 are clearly visible in the early phases of
evolution (around $2$~Gyr, Fig. \ref{fig:fdenmap}). The evolution of
E-05 is more similar to the model with the non-rotating halo (A+00).

In addition to a bar mode, the stellar discs in these models also
develop higher order modes for which we restrict our analysis only to
the $m=4$ component. In Fig.~\ref{fig:A4ps}, we show the time
evolution of the $m=4$ Fourier component of the disc surface
density. It is apparent from the figure that $m=4$ mode is significant 
in model D+07, the one with the highest halo spin in our sample. Not
only a bar but also such an $m=4$ component would take part in taking
away angular momentum from the stellar disc and thereby accelerate the
overall transformation into a bar. The stellar discs of all $5$ models
do not grow any significant $m=1$ mode (or higher-order odd modes).

\subsection{Bar size and halo spin}
\label{sec:size}
After $\sim 3$~Gyr, all the models have grown a bar in their stellar
discs. The size of a bar increases as time progresses by loosing 
angular momentum primarily to the dark matter halo. We compute the bar size 
in our model galaxies to understand its variation with the dark matter halo spin.
There exists a number of ways to measure the size of a bar in a galaxy 
\citep{Erwin2005}. In N-body simulations, the bar size measurements can be obtained
from the radial profile of the $m=2$ Fourier component ($A_2(r)$) and its phase 
angle variation with radius. We take the full width at half maximum (FWHM) of $A_2(r)$
as the first measurement of a bar size. Theoretically, the phase angle of a bar 
does not vary with radius but in reality, they do. We consider the radius 
at which the phase angle varies by $\sim 5^{\circ}$ as the second measurement 
of the bar size. We take
the mean of these two measurements to specify the bar size in our simulations.
In Fig.~\ref{fig:barsize}, we show the bar size for the different
models at two different epochs (after 3 Gyrs and 5 Gyrs). It is very
clear that at a given time the bar has - at all epochs - significantly
larger sizes for halos with larger spin parameters. The trend seems to
be even non-linear. Note that in the case of a fast co-rotating halo,
the bar size has increased by $\sim 50\%$ in $2$~Gyrs. 

\subsection{Halo spin and the Ostriker-Peebles criterion}
\label{sec:OPspin}
\cite{Ostriker-Peebles1973} showed how an otherwise bar unstable
stellar disc becomes more stable against bar formation as the mass of
a surrounding dark halo increases. We show the time evolution of
$t_{\mathrm{OP}}$ for our models in Fig.~\ref{fig:top} (similar to
Fig.~6 of \cite{Ostriker-Peebles1973}). The initial   
value of $t_{\mathrm{OP}}$ for the model A+00 suggests that the disc should be
stable against bar formation. Increasing the value of the halo 
spin (keeping its mass constant) makes the system more bar unstable
and model D+07 has $t_{\mathrm{OP}}=0.115$ and therefore should be marginally bar
unstable. Indeed, this model develops the strongest early bar in our
simulations which can be interpreted in the sense that the spin of
the co-rotating halo increases the net rotational motion in the
galaxy as a whole. However, other models (B+03 and C+05) also quickly
develop bars and from our analysis we find that a hot stellar disc
can become bar unstable within a reasonable time if $t_{OP}>0.03$ which
also happens to be the most probable spin parameter for a $\Lambda$CDM
halo. However, in order to properly understand bar formation in
galaxies, one needs to take into account the combined effect of the
mass and spin (angular momentum) of the dark matter halo.   

\begin{figure}
\rotatebox{0}{\includegraphics[height=10.5cm]{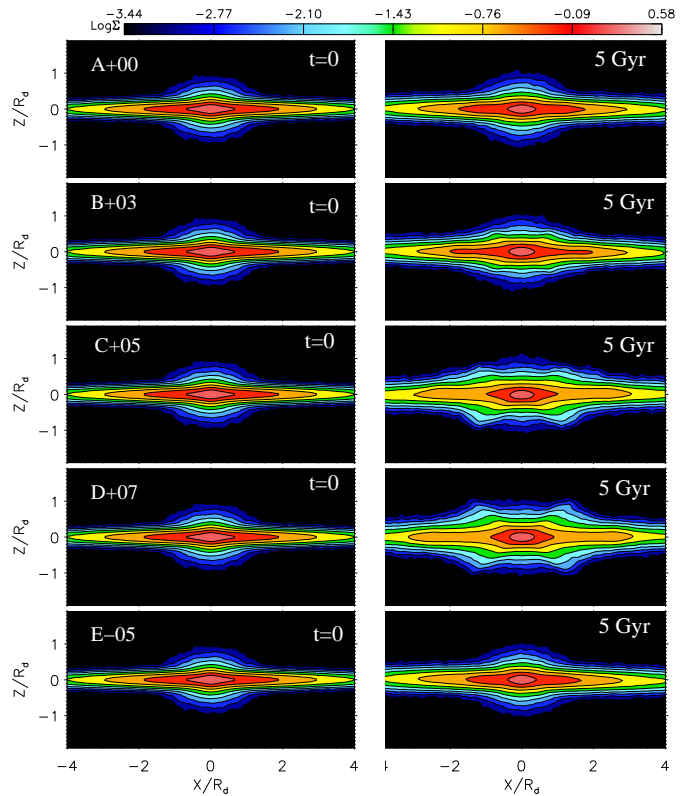}}
\caption{Edge-on galaxy morphology (stellar surface density)
  after 5 Gyrs of evolution (right) compared to the initial
  conditions (left). From top to bottom the spin of the dark matter
  halo increases; except for the bottom panel (counter-rotating halo
  model E-05). Spinning co-rotating dark matter halos promote the
  formation of peanut-shaped/boxy bulges (i.e. D+07).}  
\label{fig:edgdenmap}
\end{figure}

\begin{figure}
\rotatebox{270}{\includegraphics[height=7.5 cm]{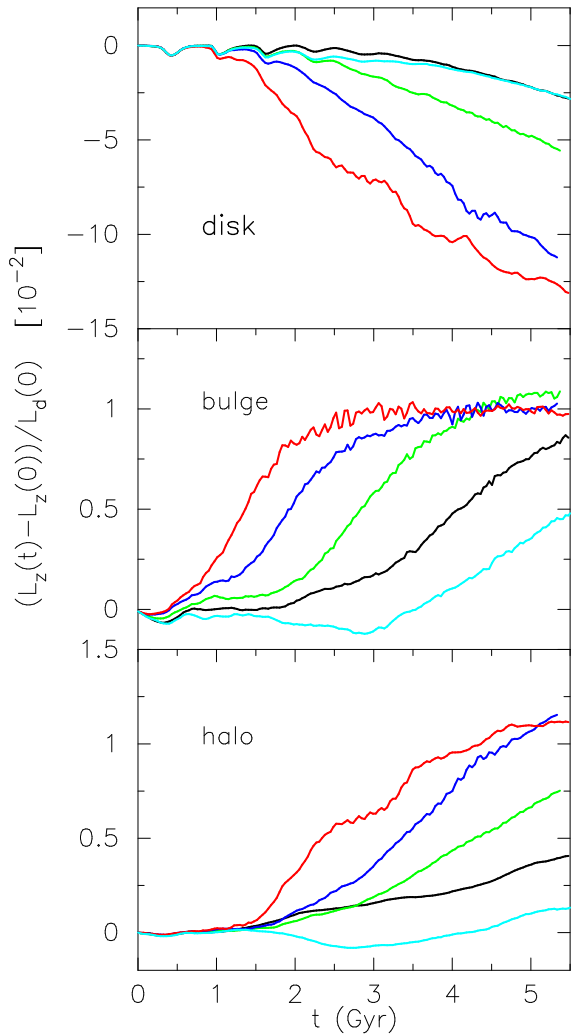}}
\caption{Time evolution of specific angular momentum of disc stars
  (upper panel), bulge stars (middle panel) and halo particles (bottom
  panel) for all 5 galaxy models. $L_d(0)$ denotes the disc specific
  angular momentum at $t=0$. The red line denotes the model D+07, 
blue C+05, green B+03, black A+00, and cyan E-05. The mass of the halo 
is $57$ times  higher than that of the bulge. As expected the model developing the
  stronger bar transfers most angular momentum (D+07).}
\label{fig:spLz}
\end{figure}

\begin{figure}
\rotatebox{270}{\includegraphics[height=8.5 cm]{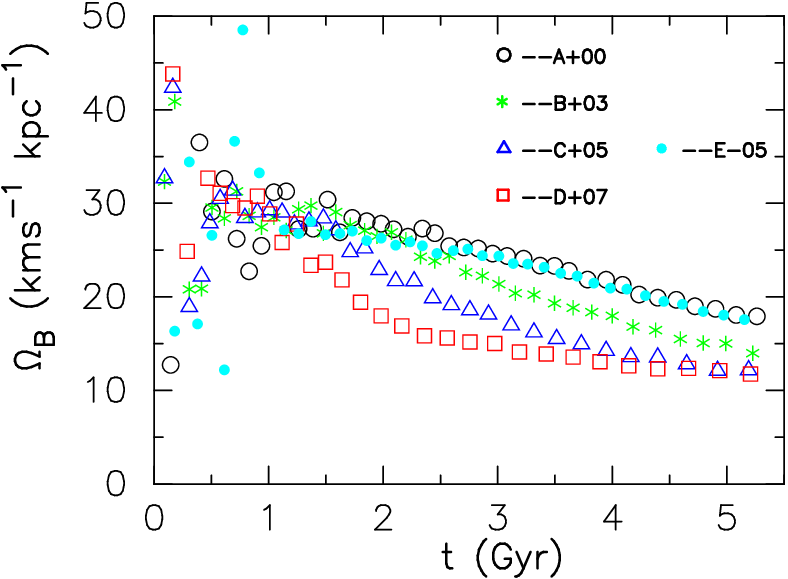}}
\caption{Time evolution of the pattern speeds of the bars,
  $\Omega_{\mathrm{B}}$, for all models. In models with more rapidly
  spinning halos the bar pattern speed is lower at late times.}  
\label{fig:pattspd}
\end{figure}

\subsection{Bulge formation}
\label{sec:bulge}
Numerous N-body simulations have demonstrated that kinematically cold
and thin stellar discs easily form bars that can transform into a
central bulge-like structure, sometimes termed pseudo-bulges 
\citep[see][for a review]{KormendyKennicut2004}. Boxy/peanut-shaped
bulges are a class of bulges primarily formed via the buckling (or
fire-hose) instability of a strong bar \citep{Combesetal1990,Rahaetal1991,
HernquistWeinberg1992,Athanamisi2002,Debattistaetal2004,KormendyKennicut2004,
  Sahaetal2010,Sahaetal2013}. A detailed orbital analysis by 
\cite{PfennigerFriedli1991, Patsisetal2002} shows that the presence of 
$2:1$ and $4:1$ vertical resonance plays an important role in the formation of 
such a boxy/peanut bulge.   
The time-scale to build such a bulge-like component in an isolated
disc galaxy depends on various parameters such as the dark matter
content, the value of Toomre Q in the disc, and might as well be
influenced by a pre-existing 'classical' bulge
\citep{Sahaetal2012}. For kinematically cold discs ($Q \sim 1 $),
building a pseudo-bulge can take only one Gyr or less. In hot ($Q > 2$) 
and dark matter dominated discs, the time-scale to form a pseudo-bulge
can be very long; in some cases it might even take a Hubble time
\citep{Sahaetal2013}. 

\begin{figure*}
\rotatebox{0}{\includegraphics[height=22cm]{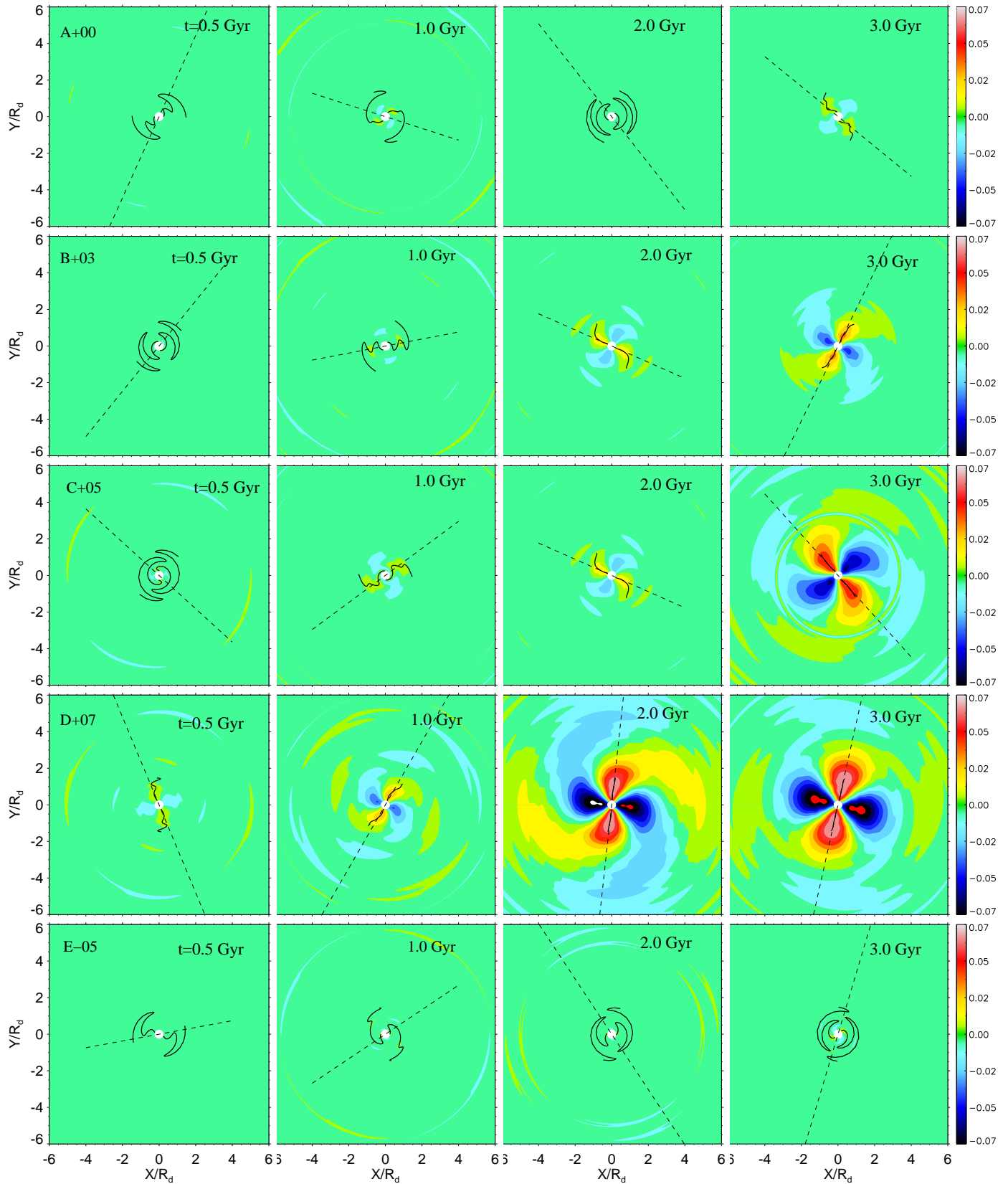}}
\caption{Time evolution (from left to right) of dark matter density
  wakes in the equatorial plane of the discs for models with varying
  halo rotation (from top to bottom). Red (blue) indicates over-dens
  (under-dens) regions with respect to the mean density at a given
  radius. The dashed line indicates the position angle of the stellar
  bar and the solid line follows the peak overdensity in the
  halo. The model with the highest halo spin (D+07) develops the
  strongest trailing density wakes in the halo (e.g., fourth row,
  second and third panel from the left). Trailing wakes result in
  angular momentum loss by dynamical friction.}    
\label{fig:wakemap}
\end{figure*}

In Fig.~\ref{fig:edgdenmap}, we compare the final ($T = 5$~Gyrs)
edge-on stellar morphologies (projected stellar surface density) of
all galaxy models to the initial conditions ($T = 0$). The models with
a non-rotating halo (A+00) and a counter-rotating halo (E-05) show
bulge structures similar to the initial conditions, except for some
amount of thickening due to heating by the weak bar
\citep{Sahaetal2010}. Although the stellar discs in both models (A+00
and E-05) form moderately strong bars at the end, they did not go
through a buckling instability phase and have not developed a
boxy/peanut-shaped bulge; see  \cite{Sahaetal2013} for a similar case
in a dark matter dominated galaxy model. On the other hand, all models
with a co-rotating halo develop bars that are strong enough to develop
a buckling instability and form peanut-shaped/boxy bulges. 
The effect is strongest for the model with the highest dark
matter spin (D+07). The buckling instability in this model occurred at around 
$3$~Gyr - a corresponding drop in $A_2$ was not apparent from the lower panel 
of Fig.~\ref{fig:A2ps} as has been suggested in many previous works 
\citep{Athanassoula2012}. But a mild drop in $A_2$ was seen when we used 
integrated $A_2$ (see upper panel of Fig.~\ref{fig:A2ps}). Note that in a 
recent work by \cite{Sahaetal2013}, it has been suggested that the formation epoch
of the boxy/peanut bulge is better correlated with the tilt of the velocity ellipsoid.   
The formation of boxy/peanut bulges not necessarily 
requires a fast and strong bar; they can form out of the slowly rotating
bars as we have shown here. We compute the relative extent of the peanut length (see
Fig.~\ref{fig:edgdenmap}) with respect to the bar for the model D+07 following 
\cite{SahaGerhard2013b}. We found that at $z = 0.4 R_d$, the length of the peanut 
for this model at $5$~Gyr is $\sim 60$\% of the bar size (see Table~\ref{tab:paratab}). 
The orbital support for such a peanut bulge to form
from a slowly rotating bars was found explicitly by \cite{Skokosetal2002}
and their study could be relevant for a detailed understanding of the peanut
bulges that form in our simulations (also for models A+00 and E-05 which did not form
a peanut bulge).    
As the galaxy models in our simulations host pre-existing classical
bulges, the models with halo spin develop composite bulges with a
different origin and morphology. We emphasize here that knowing the mass model alone
is not enough to understand the formation of a boxy/peanut bulge or composite bulge, 
one needs to have an access to the spin of the dark matter halo.  

\section{Transfer of Angular momentum and dynamical friction}
\label{sec:TAM}
Once a bar starts to form from disc stars, it grows via non-linear
gravitational interactions involving the exchange of angular momentum
with the outer parts of the disc, the surrounding dark matter halo
and a pre-existing classical bulge (if present). The angular momentum
is transferred by the gravitational torque produced by the bar
\citep{Lynden-BellKalnajs1972}. Thereafter, it has been emphasized by
several authors \citep{TremaineWeinberg1984, Weinberg1985,
  HernquistWeinberg1992, WeinbergKatz2002,Athana2002,
  Athanassoula2003,SellwoodDebattista2006,Dubinskietal2009,Sahaetal2012}
that orbital resonances between disc stars and the particles
representing the dark matter halos play a significant role in the
transfer of angular momentum between the bar and the rest of the
galaxy, with dark matter halo particles taking away most of the
angular momentum from disc stars. In section~\ref{sec:darkbar} we
discuss, in detail, the physical process that allows the dark halo to
do so. Most previous studies have investigated the transfer of angular
momentum to a non-rotating dark matter halo. Here we quantify the
angular momentum exchange between a bar and a rotating dark matter
halo in more detail.   

In Fig.~\ref{fig:spLz}, we show the time evolution of the specific
angular momentum of the disc, bulge, and dark matter halo for all
models. As mentioned above, the exchange of angular momentum between 
the stars and dark matter halo occurs when 
a resonant condition $l\kappa + n\Omega - m \Omega_B=0$ between the bar and 
halo orbit is satisfied, $\Omega_B$ being the bar pattern speed. For the 
bar mode ($m=2$), the important resonances are in this context 
$(l,n,m)=(-1,2,2)$ i.e., ILR; $(l,n,m)=(0,2,2)$ i.e., co-rotation.
Independent of halo rotation, all stellar discs lose angular
momentum which is deposited in the stellar bulge and the dark matter
halo. The rate of angular momentum transfer correlates with the spin
of the dark matter halo (or the strength of the bar). 
The amount of angular momentum transferred is highest in model D+07,
qualitatively, in accordance with the semi-analytic results from
\cite{Weinberg1985}. According to his calculation, the rate of
angular momentum loss from the bar 
monotonically increases as the fraction of prograde orbits in the 
halo increases from $0$ to $1$. The reason for this is that prograde
orbits contribute predominantly to the $(-1,2,2), (0,2,2)$
  resonances which are, in general, stronger than $(-1,-2,2),
  (0,-2,2)$ resonances for a given $l$. For the same reason, one
would expect less angular momentum transfer to a halo with a large
fraction of retrograde orbits. In model E-05 with a counter-rotating
halo, we see that the angular momentum loss from the stellar disc is
nearly negligible up to $\sim 3$~Gyrs and only after 
$4$~Gyrs it is comparable to the case of a non-rotating
halo. Interestingly, the counter-rotating halo looses angular
momentum between $\sim 1.5 - 4$~Gyrs of evolution. Only for model D+07
there is an indication that the angular momentum transfer to the halo
might saturate.  

As a result of the angular momentum loss from disc stars, the pattern
speed of the bars decreases (Fig.~\ref{fig:pattspd}). Models with
stronger bars (more rapidly spinning dark matter halos) show a more
rapid slowdown of the bar in accordance with \citep{Weinberg1985}. 
The effect of halo rotation on the dynamical friction and hence the 
slow down of bar pattern speed was previously studied by 
\cite{Athanassoula1996, DebattistaSellwood2000} while the former found 
the bar pattern speed to be decreasing rather slowly in a co-rotating
halo. The bar in the model with the fastest
rotating halo (D+07) slows down faster during the initial phase of
evolution; the pattern speed of the bar in D+07 is nearly half of that
in A+00 (non-rotating halo) at $\sim 2$~Gyr. Non-linear effects are
also apparent from  this figure. In model D+07, the pattern speed of
the bar remains nearly unchanged from $\sim 2.5$~Gyr onwards although
the halo keeps absorbing angular momentum (see Fig.~\ref{fig:spLz}). 
Since the bar in this model keeps growing, we think most of the
angular momentum loss is likely to be invested into changing the moment of inertia
of the bar according to:

\begin{eqnarray}
\frac{dL_z}{dt} = I_B \frac{d\Omega_B}{dt} + \Omega_B \frac{dI_B}{dt},
\end{eqnarray}

where $I_B$ and $\Omega_B$ are the moment-of-inertia and pattern speed
of the bar. 
Notice also that the angular momentum gain by the pre-existing
classical bulge (in model D+07) nearly saturates beyond $\sim 2.5$~Gyr - a
fact that correlates well with bar pattern speed slow down beyond
$\sim 2.5$~Gyr in the same model. The pre-existing classical bulge
is constructed with a streaming fraction of $0.5$ i.e., half of the 
bulge stars are in retrograde orbits. Fig.~\ref{fig:spLz} shows that
the bulge stars initially looses angular momentum as in the counter-rotating 
halo model. Also all the bulges seem to be absorbing angular momentum earlier than the halo
in our models. A detailed analysis of these findings is beyond the scope and aim of this paper
and will be followed up using orbital analysis in a future paper. 
Overall, we find that a co-rotating halo with a reasonable spin parameter is capable of
absorbing a significant amount of angular momentum from the disc (due
to the formation of a stronger bar). As a result the bar can be slowed
down by factor of $2$ within about a Gyr, i.e. about $27$ dynamical 
time scale computed at the half-mass radius after its formation. 

\subsection{Density wakes and the dark bar}
\label{sec:darkbar}
A rotating self-gravitating bar inside a live dark matter halo will
create a density wake in the halo density distribution that in turn
exerts a torque on the bar and slows it down. This phenomenon is known
as dynamical friction \citep{Chandra1943,TremaineWeinberg1984} and has 
been  studied, in great detail, by several authors, especially in the 
context of bar-halo friction \citep{Weinberg1985,DebattistaSellwood2000, 
Holley-Bockelmannetal2005, SellwoodDebattista2006, WeinbergKatz2007a}. 
Previous investigations of the bar-halo friction
using N-body simulations, however, considered only non-rotating
halos. In the following, we study the nature of density wakes in models
with rotating halos in more detail. 

\begin{figure*}
\rotatebox{0}{\includegraphics[height=4.4 cm]{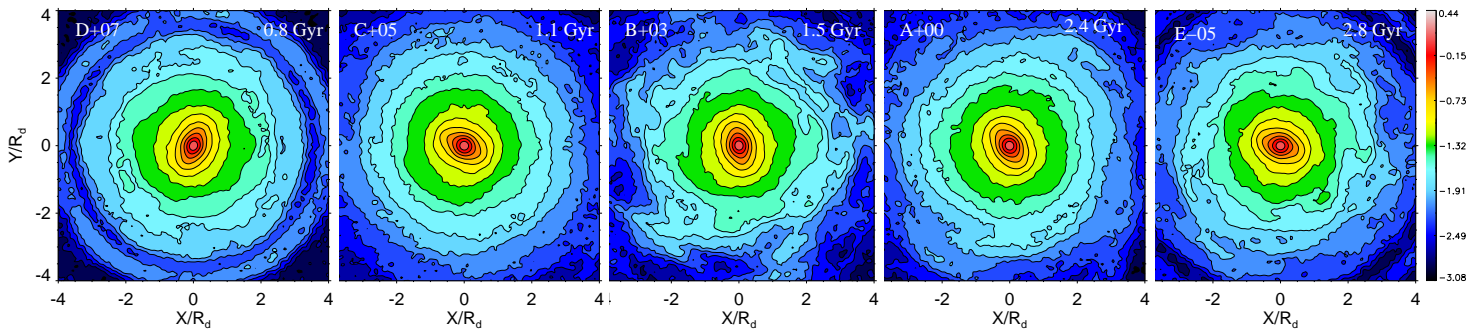}}
\caption{Surface density of stars of each model galaxy shown at the
  time when the bar has just formed. Formation times are given in the
  upper right corner of each panel. The bar forms first (after 0.8
  Gyrs) in the model with the highest halo spin (D+07, left) and
  last (after 2.8 Gyrs) in the model with the counter-rotating halo
  (F-05,right).}
\label{fig:denA2-0.2}
\end{figure*}

\begin{figure}
\rotatebox{270}{\includegraphics[height=8.5 cm]{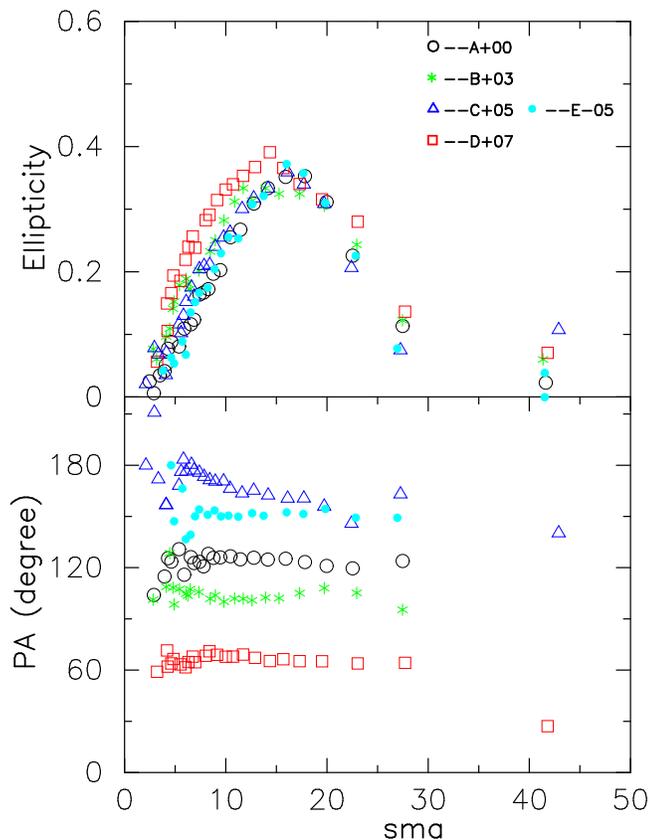}}
\caption{Ellipticity and position angles (PA) as a function of the
  semi-major axis of the bars at the time of their formation (see
  Fig.~\ref{fig:denA2-0.2}). All bars have similar ellipticity profiles
  but different position angles and pattern speeds (see
  Fig.~\ref{fig:pattspd}). One sma unit corresponds to $0.04 R_d$.}
\label{fig:ellipPA}
\end{figure}

We decompose the density distribution of the dark matter halos in different models
 into spherical harmonics
$\rho_h(r,\theta,\varphi)=\rho_{lm}(r)Y_{lm}(\theta,\varphi)$. The
density component $\rho_{22}$ corresponding to $l=2,m=2$ is the
strongest, although the contributions from $l=4,m=2$ and $l=4,m=4$ are
non-negligible, especially in models D+07 and C+05. In
Fig.~\ref{fig:wakemap}, we show the density distribution $\rho_{22}$  
projected onto the equatorial plane ($\theta=90\,^{\circ}$) for each model at 
different times. In model A+00 (top
row), we see a tiny bar-like density wake produced at about
$3$~Gyr. By the same time the density distribution of the
counter-rotating halo model (E-05) remains nearly featureless without
any density wakes . In contrast, strong bar-like density wakes are
produced in all models with co-rotating halos; with increasing
intensity as the spin parameter increases. The nature of the density
wakes can be followed clearly in model D+07 (fourth row in
Fig. \ref{fig:wakemap}). In the initial phase (less than $\sim
2$~Gyr), dark matter particles corresponding to the $l=2,m=2$ density
wakes form a spiral-like structure which lags behind the bar in the
stellar disc. The phase angle of the spiral-like structure with
respect to the stellar bar varies with radius. As time progresses, the
density wakes get stronger and forms a bar-like feature in the inner
part with a nearly constant phase out to $\sim$ 2 disc scale
lengths. At larger radii the wakes still have a spiral-like structure
and lag behind the stellar bar. The swarm of dark matter particles
in the spiral-like wakes  lagging  behind  the stellar bar are the
primary source of dynamical friction. They exert torques on the  
bar and slow it down. At around $3$ Gyr
the spiral-like structure in model D+07 disappears, leaving behind 
a bar-like structure which is almost in co-rotation with the stellar
bar. This feature has been called a dark bar or halo bar in the 
literature \citep{Holley-Bockelmannetal2005,Athanassoula2007}. This 
dark bar is a resonance driven feature and would be hard
to distinguish from the stellar bar as it has nearly the same pattern
speed. The same process happens in all other models with co-rotating
halos, however less rapidly for halos with lower spin parameters.

\section{Resonances and slow bars}
\label{sec:reso}
To understand the nature and dynamics of the bars forming in models
with spinning halos, we identify the birth time of the bars by
inspecting the maximum values of $A_2/A_0$, the radial variation of
the projected ellipticity and the position angle of the semi-major
axis of the bars. As explained above, dark matter halos with high spin
parameter promote the early formation of bars, which is quantified
here. In Fig.~\ref{fig:denA2-0.2}, we show the face-on stellar surface 
density maps for all models at the corresponding birth time of the
bars which we have defined in the following way: a bar is 'born' when
the peak ellipticity ($\epsilon_{\mathrm{max}}$) along  the bar
semi-major axis becomes larger than $0.3$, and the the corresponding
$PA$ is almost constant. In Fig.~\ref{fig:ellipPA}, we show the
ellipticity and the position angle as a function of the semi-major
axis (sma). The aforementioned range of the peak ellipticity roughly
corresponds to $(A_2/A_0)_{max} \sim 0.2$ (see Fig.~\ref{fig:A2ps}).     
At formation, all bars have very similar shapes but they have
different pattern speeds. We compute the size of the nascent bar using
only the radial variation of its PA in the same way as explained in
section~\ref{sec:size}. This way, we get the largest size of the bar,
called as $L_{bar}$ in the literature (e.g. \citealp{Erwin2005}), in
all models and the corresponding values are given in
Table~\ref{tab:paratab2}.  We use the estimate of $L_{bar}$ to get a
lower limit on the ratio of the co-rotation radius to the radial
extent of the bar, $R_{cr}/R_{bar}$,  which decides whether these
initial bars are slow or fast in nature. A bar is termed 'fast' if  
$1.0 < R_{cr}/R_{bar} < 1.4$ and 'slow' for which the ratio is greater
than $1.4$  \citep{Aguerrietal2003}.

\begin{table}
\caption[ ]{Properties of nascent bars. The unit of pattern speed $\Omega_B$ is in 
km s$^{-1}$kpc$^{-1}$.}
\begin{flushleft}
\begin{tabular}{ccccc}  \hline\hline 
Models      & $L_{\mathrm{bar}}/R_d$ & $\Omega_{\mathrm{B}}$ & $R_{\mathrm{cr}}/R_d$ & $R_{\mathrm{cr}}/L_{\mathrm{bar}}$ \\
       & &  &   &\\
\hline
\hline
A\Plus00   & 0.85 & 26.9 & 2.20  & 2.58 \\
B\Plus03   & 0.90 & 27.0 & 2.18  &2.42\\
C\Plus05   & 0.72 & 29.0 & 2.00  &2.77 \\
D\Plus07   & 1.10 & 29.5 & 1.95  &1.77\\
E\Minus05   &0.80 & 24.5 &2.50   & 3.12\\

\hline
\end{tabular}
\end{flushleft}
\footnote{XYZ}
{$L_{\mathrm{bar}}$ is the length of the bar measured based on the phase angle variation. Rest of the symbols are explained in Table~\ref{tab:paratab}}.
\label{tab:paratab2}
\end{table}

\begin{figure}
\rotatebox{270}{\includegraphics[height=8.5 cm]{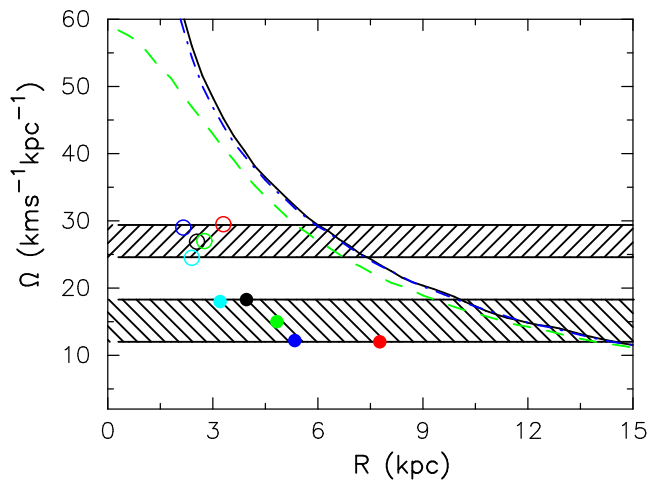}}
\caption{Location of co-rotation resonances in the stellar disc which 
has just formed a bar and at a later time, $5$~Gyr. The forward hatched
region denotes the range of pattern speeds amongst the nascent bars
(at the time of their formation, see Fig. \ref{fig:denA2-0.2}) and
the back-hatched region for the evolved bars at $5$~Gyr. The ranges of 
pattern speeds are given in Table~\ref{tab:paratab} and Table~\ref{tab:paratab2}. 
The solid black and dashed green curves refer to $\Omega$ for D+07 at those two
instances and dash-dot curve for the model A+00 at $5$~Gyr.
Open circles refer to the sizes and pattern speeds of the nascent bars while
filled circles are those of the evolved bars. The colors are identical
to previous figures.}
\label{fig:reso}
\end{figure}

\medskip
Next, we calculate the locations of the resonances in the stellar disc
to see whether spinning halos preferentially create slow bars. 
Observations and simulations (without halo rotation) by
\citet{Chemin2009} already indicated that systems dominated by dark
matter can only develop slow bars. Fig.~\ref{fig:reso} shows the location
of co-rotation resonances when the bar has just formed in the disc and towards
the end of the simulation. We have checked that there was no ILR in the stellar 
disc initially. From Fig.~\ref{fig:pattspd}, we know that the nascent bar in model D+07
rotates faster than all others (open red circle in
Fig. \ref{fig:reso}) with CR at $\sim 6$~kpc. The slowest rotating
bar in model E-05 (couter-rotating halo, open cyan circle)
pushes CR to $\sim 7.5$~kpc. Comparing, the bar sizes and CR
radii we find that  $1.77 < R_{cr}/R_{bar} <3.1$  which is larger than
$1.4$ for all models, i.e. above the limit for fast bars. This
property does not change with time. After $5$~Gyrs of evolution the
bars are in the range $1.83 < R_{cr}/R_{bar} < 3.1$. The values for bar
sizes, CR radii and pattern speeds after $5$~Gyrs are quoted in
Table~\ref{tab:paratab}.    

We conclude that all bars in the galaxy models presented here form as
slow bars and they remain slow bars as the galaxy evolves. The
transformation of a slow bar to fast is unlikely.  Increasing the spin
of a halo (to more unlikely values) could possibly reduce the ratio
$R_{cr}/R_{bar}$ close to $1.4$ leading to the possibility of
harboring a fast bar in a dark matter dominated galaxy. It is
worthwhile to point out that bars in early-type galaxies are generally
fast, while they could either be slow or fast in late-type galaxies
\citep{Aguerrietal2003,Rautiainenetal2005}.    

\section{Summary and discussion}
\label{sec:discuss}

In this paper we have demonstrated that spinning dark matter halos
promote the formation of stellar bars, the most common and most
intriguing non-axisymmetric feature in galaxies. Even dark matter
dominated hot stellar discs, which are expected to be stable 
against bar formation according to traditional criteria 
\cite[see e.g.,][however, see
  \citealp{Athanassoula2008}]{Ostriker-Peebles1973,ELN1982} ,  
can form strong bars depending on the spin of their co-rotating host
halos. Stellar discs embedded in faster spinning halos - we have
tested halos in the range $0 <  \lambda_{\mathrm{dm}} < 0.07$ -
develop their bars up to $2$~Gyrs earlier with a higher ($m = 2$)
Fourier amplitude (up to peak $A_2/A_0 \sim 0.7$) and the final bars are
longer (up to a factor of $2.5$). The essence of effect has been suggested by 
\citep{Weinberg1985} and is caused by more efficient angular 
momentum transfer from the disc to the spinning dark matter halos due
to resonant interactions between disc stars and co-rotating halo 
particles. We have demonstrated that dynamical friction between a
bar and a halo is stronger when the halo is co-rotating with the disc
leading to significant transfer of angular momentum to the dark matter
particles which form a bar-like structure in the inner region. 
As a result, a higher local density of dark matter particles can be 
expected wherever the stellar bar is located. This effect is weakest 
in counter-rotating halos (we have tested one model) which thus appear 
to suppress the formation of bars. The presence of counter-rotating
halos could plausibly explain why some of the observed massive and
cold stellar discs did not form any bar at all  \citep{Shethetal2012}.     

The dynamical friction between a rotating bar and spinning halo requires 
that the halo is composed of particles not just a rotating potential.
Hence the effect would be visible when the dark matter halo is live
i.e., is able to interact with the disk stars. In other words, rigid
halos might not show the same dramatic effect.

Not all bars in our models form boxy/peanut-shaped (pseudo-) bulges in
the end \citep{KormendyKennicut2004,Athanassoula2005}. Although the
$m=2$ peak amplitudes of all the bars are nearly the same after
$5$~Gyr, only the discs embedded in co-rotating halos with spin
parameters $\lambda_{\mathrm{dm}} > 0.03$ went through a buckling
instability phase. Discs embedded in faster spinning halos develop
stronger bars and appear more box-shaped in the end.   

The impact of halo spin on bar formation in the idealized model discs 
discussed here is unambiguous and might have important consequences
for the formation and evolution of disc galaxies in $\Lambda$CDM
cosmologies. As discussed in section~\ref{sec:introduc}, the rotation
of dark matter halos is a definite prediction from large scale dark
matter simulations with $\Lambda$CDM cosmological models
\citep{Steinmetz1995,Bullocketal2001,Vitvitskaetal2002,Springeletal2005, 
Bettetal2007}. As most of the stellar discs are expected to co-rotate 
with their host halos \citep[see e.g.,][]{Salesetal2012,Aumerwhite2013}, 
the formation of bars might be promoted in general. It has been shown by a number of
studies that systems with already massive and dynamically cold stellar
discs become bar unstable easily \citep{CombesSanders1981,Valenzuela-Klypin2003,
Dubinskietal2009,Sellwood-Debattista2009,Sahaetal2012,Athanassoulaetal2013} 
and a rotating halo might only accelerate the instability in those systems. 

More interesting is the effect on systems which traditionally were not
expected to be bar unstable but are observed to have bars. In
particular, present day low-mass galaxies which are, in general,
observed to be dark matter dominated
\citep{Barazzaetal2008,Shethetal2008, Chemin2009,NairAbraham2010} and
- in the $\Lambda$CDM framework - are expected to have converted only
a very small fraction of the available baryons into stars
\citep{Yangetal2012,Mosteretal2013,Behroozietal2012}. Another class of
galaxies for which the above effect might be very relevant are low
surface brightness galaxies (LSB). They are dark matter dominated,
their stellar discs are relatively hot at a very low level of star
formation despite having a similar amount of gas as in their high
surface brightness (HSB) counterparts. LSB galaxies are generally
found in isolated environments which excludes the possibility of
strong tidal interaction yet bars  are found in these galaxies
\citep{MatthewsGallagher1997,Pohlenetal2003}. Previous  N-body
simulations by \citep{MayerWadsley2004} showed that LSB stellar discs
embedded in CDM halos are generally stable against bar formation. Our
study with spinning halos sheds new light on the prospect of bar
formation in LSB galaxies.  

Another interesting (but definitely more speculative) regime for this process to
become important is at high redshift ($z >  2$) where star forming
galaxies have hot disc components \citep{ForsterSchreiberetal2009} and a
significant fraction  of the baryons is still in the form of gas
\citep{Taconietal2012}. Here rotating halos might promote the early
and rapid formation of bars with important consequences for
instability driven bulge evolution models \citep{Genzeletal2008} and,
eventually, the formation and early growth of supermassive black holes 
\citep{Boweretal2006,Fanidakisetal2012,Hirschmannetal2012}.      

\label{lastpage}

\section*{Acknowledgement}
\noindent We thank Jerry Ostriker, Lia Athanassoula, Panos Patsis for valuable 
comments on the manuscript. We thank the anonymous referee for useful comments on 
the draft and especially for suggesting to use integrated 
bar amplitude. K.S. acknowledges support from the Alexander von Humboldt
Foundation during which this project started.

\end{document}